\documentclass[%
 reprint,
 amsmath,amssymb,
 aps,
 pra,]{revtex4-2}
 
\usepackage{hyperref}
\usepackage{graphicx}
\usepackage{dcolumn}
\usepackage{bm}
\usepackage{color}

\begin{document}
\preprint{APS/123-QED}
\title{Super-Bloch oscillations with parametric modulation of a parabolic trap}
\author{Usman Ali}
\author{Torsten Meier}%
\affiliation{%
 Paderborn University, Department of Physics,\\
 Warburger Strasse 100, D-33098 Paderborn, Germany
}%

\date{\today}
       
\begin{abstract}
Super-Bloch oscillations are the outcome of a relative phase between Bloch oscillations and modulations of the periodic lattice. We analyze the dynamics for a model system in which such a relative phase is intrinsically present due to the position-dependent force provided by a parabolic trap and therefore an external detuning is not required. The relative phase is not unique and the realized dynamics depends on the initial phase of the modulated parabolic potential. We provide accurate explanations for the different obtained oscillatory transport and spreading regimes by analyzing the spatio-temporal dynamics in real space and by visualizing the relative phase in the k-space dynamics. We also compare our numerical results to an approximate semiclassical analytical expression for the group velocity for a modulated constant force system and find good agreement for coherent oscillations but deviations for oscillations with spreading dynamics which altogether supports the interpretations of our findings.
\end{abstract}

\maketitle

\section{\label{sec:level1}Introduction}

\par{Ultracold atomic gases in optical lattices are ideal model systems for quantum simulations \cite{1}. These systems allow for a flexible manipulation of the parameters, inter-atomic interactions, and lattice defects. Improved measurement techniques opens doors to new perspectives, effects, and applications \cite{1}. One of the most significant example are Bloch oscillations (BO) which are largely explored in various ultracold atomic ensembles, such as degenerate Bose/Fermi gases \cite{2,3}, strongly correlated atoms \cite{4,5}, and Bose-Einstein condensates \cite{6,7,8,9}. In BOs the width or mean position of a localized wave interacting with a periodic lattice and a linear potential oscillates periodically due to periodic oscillations of the quasimomentum \cite{10,11,12,13,14,15}. The effect of the linear potential is often described in terms of a constant force which causes the quasimomentum to increase linearly with time while the lattice provides periodic bounds of the first Brillouin zone (BZ) giving rise to Bragg reflections.} 

\par{An additional advantage offered by ultracold atomic systems is the access to selective modulation of the optical potential or the external force \cite{1,16}. The later imposes a shaking effect on the lattice which for a purely oscillating drive, induces rapid oscillations of the quasimomentum. Accordingly, the wavepacket either performs slow oscillatory spreading or undergoes coherent directed transport depending upon the initial phase of the drive \cite{17,18}. It has also been demonstrated that the spreading and transport dies out completely, at specific values of the ratio between the amplitude of the driving field and its frequency. This leads to a regime of the coherent destruction of tunneling, also termed as dynamical localization \cite{19}.}

\par{The spreading and transport dynamics also exist in the combination of BOs for a modulation of the constant force \cite{20,21}. Here they occur at particular resonances at which the driving frequency is an integral multiple of the Bloch frequency. The modulation leads to additional small oscillations of the quasimomentum, which appears at specific k-values depending upon the initial phase of the drive, on top of its linear increase. For an equal distribution of the momentum around the center of the BZ this gives the ballistic spreading of the wavepacket, while in the opposite case a net transport per Bloch cycle is generated \cite{23}. At a non-resonant driving, that is a constant detuning between the driving frequency and Bloch frequency, the oscillatory modulation of quasimomentum attains a relative phase which covers the entire BZ. As a result the wavepacket transport changes its direction with the changing sign of the phase bringing slow oscillatory transport of large real space amplitude, known as super Bloch oscillation (SBO). The amplitude and the period of these oscillations is mainly dependent upon the detuning and they are inversely related to each other. These dynamics have been observed in cold atomic systems \cite{21} and have been analyzed in related theoretical studies \cite{22,23,24,25}.}

\par{SBOs have mainly been studied by modulating the otherwise constant force. However, BOs have been predicted for particularly different classes of systems which considers the gradient of a parabolic potential \cite{26,27,28,29,30} or even higher-order gradients \cite{31}. They were also observed for a noninteracting BEC placed in an optical lattice in addition to a parabolic trap potential \cite{8}. In such a system BOs exist for a wavepacket placed away from the trap center and are characterized by an intrinsic dephasing \cite{26}. In this work, we demonstrate the existence of SBOs appearing due to a periodic modulation of the parabolic trap around its mean value. Such driving of a trap is considered in analytical studies \cite{32,33,34} and appears to be also experimentally feasible \cite{35}.}

\par{We find that by a parametric modulation of the parabolic trap potential SBOs can be generated in the absence of an external detuning. This we demonstrate by numerically calculating the spatio-temporal dynamics. Our interpretation is confirmed by monitoring the dynamics in k-space which reveal the presence of a relative phase. It is noted that in such a system the role of detuning is played by the spatial variations of the BO frequency which originates from the position-dependent force. The spatial variations are even more pronounced due to large transport and a new kind of dephasing is seen for the SBOs which has a different origin as compared to the dephasing of BOs in static systems. We find that the dephasing of the SBO can be suppressed with an initial phase shift in the driving field which can gives rise to long-lived coherent SBOs. At an opposite phase of the drive to this case we encounter asymmetric oscillatory spreading of the wavepacket.}

\par{A quick large space spreading has also observed in experiments with ${{}^7}$Li atoms where the lattice amplitude is modulated in the presence of a static trap with displaced origin \cite{36}. Here, we find that opposite to a modulation of the lattice, the modulation of the force provides an easy visualization of the modulation effects on wavepacket dynamics by monitoring the dynamics in k-space. As in \cite{36} the dynamics occur on a larger spatial scale, the phenomenon shown here would result in more pronounced losses. Therefore we use in our analysis the parameters of an experiment performed with ${{}^{87}}$Rb atoms \cite{37}, for which BOs have been shown \cite{26}. Under these conditions we trace the long time dynamics for each case and in the spreading oscillations we find new kind of mixed dynamics which exist in the combination of BO and SBO. We explain the reported dynamics from the wavepacket evolution in k-space and by comparing to an approximate semiclassical expression of the group velocity.}

\par{This paper is organized as follows: In Section~II, we introduce the model and the numerical procedure, and provide the parameters used in our calculations. We present, explain, and discuss the obtained results for SBOs in Section~III and some conclusions are provided in Section~IV.}

\section{The Model}

We consider a condensate of ultracold rubidium atoms in an axially-symmetric crossed optical dipole trap which provides loose axial confinement, as compared to tight confinement along the transverse plane. In the limit of strong transverse confinement, and considering the atom-atom interactions are tuned to zero by using the Feshbach resonance, the effective potential in the axial direction is parabolic \cite{39}. In addition a one dimensional (1D) optical lattice is introduced alongside the parabolic potential, which results a symmetrically-curved periodic potential, see Fig.~1. We assume that the dynamics starts with a rapid displacement of the center of the parabolic potential which together with its subsequent time-dependent modulation can be realized by specialized optical modulators, such as acousto-optic modulator, in the axial beam \cite{35}. Thus the parabolic potential is modulated periodically which means that the overall curvature oscillates in time. In dipole and rotating wave approximations, the dynamics of ultracold atomic condensates in the combined potential of a 1D optical lattice and a modulated parabolic trap are effectively described by the Hamiltonian
\begin{eqnarray}
 H=\frac{{P}^2}{2m}+\;V_{o}\;sin^2{\left(\frac{\pi}{d}x\right)} \quad\quad\quad\quad\quad\quad\quad\\
 +\frac{1}{2}m\omega_{\tau}^2x^2\{1+ \alpha \sin(\omega_D t+\phi)\} . \;  \nonumber
 \end{eqnarray}
Here, $V_o$ is the depth of the optical lattice, $d$ is the lattice period, $\omega_{\tau}$ is the frequency of the parabolic trap, and $m$ is the atomic mass. Moreover, $\alpha$, $\omega_D$, and $\phi$ denote amplitude, frequency, and initial phase of the driving field, i.e., the oscillatory part of the parabolic potential, respectively.

\begin{figure}[tb]
\hspace*{-0.3cm} 
\includegraphics[scale=0.28]{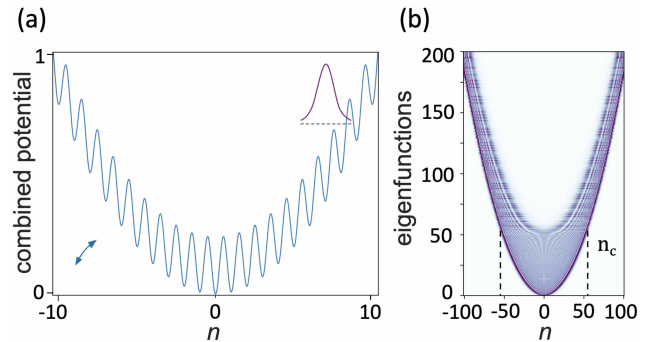}
\caption{\label{fig:epsart} (a) Schematic diagram showing the combined potential of a 1D optical lattice and an additional parabolic trap. The arrow indicates the oscillating curvature of the combined potential in the presence of a modulation of the trap and the inset represents a Gaussian wavepacket which we consider as the starting point of the quantum dynamics.
(b) Eigenfunctions of the time-independent system as a function of the real space index n. The dashed lines at $n=n_c$ mark the separation between harmonic oscillator like and spatially-localized eigenfunctions.}
\end{figure}

\par{If the lattice depth is sufficiently high as compared to the atomic recoil energy $E_R= \hbar^2\pi^2/2md^2$, i.e. $V_0\gtrsim 4E_R$, and the modulation does not induce interband tunneling, one can express the wavefunction $\psi(x,t)$, as a superposition of Wannier functions that are localized in the individual wells via $\psi(x,t)= \sum_n c_n(t) \;\!\omega_0(x-nd)$. In such a single-band tight-binding description of the system, the dynamics of the complex amplitudes $c_n(t)$ follow from the discrete time-dependent Schr\"odinger equation and reads
\begin{equation} 
	i\hbar \; \frac{\partial {c}_n }{\partial t}	= -\frac{J}{2} (c_{n+1}+c_{n-1}) + {K(t)} \; n^2 c_n ,
\end{equation}
with $K(t) = K+K_D \sin(\omega_Dt+\phi)$, the trap strength $K = m\omega_{\tau}^2d^2/2$, and the trap modulation strength $K_D = \alpha K$. The symbol $J$ denotes the nearest-neighbor tunneling matrix element, which depends upon the scaled depth of the optical lattice, $s=V_o/E_R$, as $J/E_R = {8}\;{{\left(s\right)}^{\frac{3}{4}}}e^{-2\sqrt s}/{\sqrt\pi}\;$.}

\par{In order to initiate the quantum dynamics, we start with a sudden displacement of the parabolic trap center, along the axis of the combined potential. The displacement induces a shift in the mean position of the atomic cloud and the ensemble starts its journey over the curved periodic wells of the parabolic lattice at $x_0/d=n_0$.  For a displacement above the critical index, $n_c= (2J/K)^{1/2}$ the energy of the wells is larger than the trap free bandwidth $E>4J$. In such a regime the eigenfunctions are increasingly localized in space \cite{39,40}, as shown in Fig.~1(b). Apart from the spatial inhomogeneity arising from the parabolic potential and without considering a modulation of the trap potential the expected dynamics would be quite similar to BOs in an initially static force of strength $F=2Kn_0/d$ \cite{26}. We assume a preparation of the atomic condensate in the regime of localized eigenfunctions and describe it by a displaced Gaussian wavepacket
\begin{equation}
	c_n(t=0) = \frac{1}{\sqrt{\sigma_n \sqrt{\pi}}}\; e^{-\frac{(n-n_o)^2}{2\sigma_n^2}} e^{-ik_o n} ,	
\end{equation}
with mean, initial momentum, and the spatial-width represented by $n_0$, $k_0$, and $\sigma_n$, respectively. To obtain the dynamical evolution we solve Eq.~(2) numerically by using the fourth-order Runge-Kutta method for the initial conditions given by Eq.~(3).}
\par{We use values of the system's parameters from an experiment with Bose-condensed ${{}^{87}}$Rb atoms \cite{37}, with atomic mass $m=1.1443\times10^{-25}$~kg, which are placed in an optical lattice of period $d=397.5$~nm with depth $V_0 = 12.77~E_R$ and a parabolic trap with frequency $\omega_{\tau}=2\pi\times9$~Hz. These values corresponds to $J=0.024~E_R$, $K=1.52\times10^{-5}~E_R$, and $n_c=56$. The experimental procedure matches very closely with our assumption of the wavepacket preparation, with the only difference that we chose a parabolic optical trap, instead of a magnetic trap, and that in our model we consider that the trapping potential can be modulated periodically in time. We start the dynamics from $n_0=125$, such that the initial Bloch frequency is given by $ \hbar\omega_B=2Kn_0 = 0.0038~E_R$, which corresponds to $\omega_B=2\pi\times13.76$~Hz. The trap is modulated with a strength equal to the static trap strength, i.e., $\alpha =1$, and the modulation frequency is tuned to exactly match the initial Bloch frequency, i.e., $\omega_D=\omega_B$. As is shown below, different dynamics are obtained when tuning the initial phase $\phi$
of the time-dependent trap potential which therefore acts as a control parameter.}
\par{Due to the rather low value of the trapping frequency in comparison to other experiments, the parabolic trap potential can be considered as weak.
We consider a rather wide initial wavepacket with a width in real space of $\sigma_n=3.16$ corresponding to a distance of $1.26~\mu m$.
To trace and to be able to analyze the wavepacket dynamics in k-space we evaluate the Fourier transform
\begin{equation}
		c_k(t) = \frac{1}{\sqrt{2\pi}} \int_{-\infty}^{\infty} c_n(t)\; e^{ikn} dn .
\end{equation}
The initial width in quasimomentum space is determined by the relation $\sigma_n \, \sigma_k = 1/2$. The dynamics of the average group velocity is obtained from the gradient of the mean position as function of time. For the undriven static system the eigenfunctions shown in Fig.~1(b) are obtained by the imaginary time propagation method applied to Eq.~(2).}

\begin{figure}[tb]
\hspace*{-0.3cm} 
\includegraphics[scale=0.32]{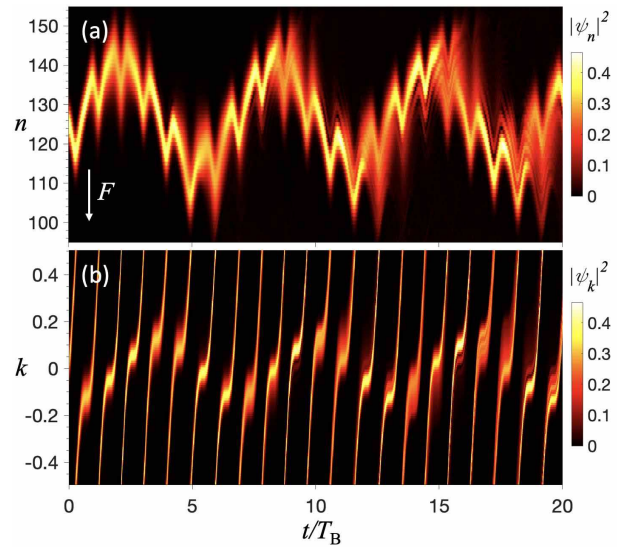}\caption{\label{fig:epsart} Time evolution of the absolute square of the wavefunction in (a) real and (b) quasimomentum space
showing SBOs with some weak additional dephasing.
The wavepacket starts its journey at $t=0$ at $n_0=125$ with $k_0=0$ and we consider a modulated trap potential with drive phase $\phi = 0$.}
\end{figure}

\section{Results and Discussions}

The dynamics obtained for our spatially-inhomogeneous and periodically-driven system for $\phi=0$ is shown in Fig.~2(a).
We find clear SBOs over a distance which is significantly larger than the localization of the initial eigenfunctions near the center of starting position $n_o$.
In more detail, a slow oscillatory transport in the combination of BO is visible which completes its period in approx. 6 Bloch periods moving across more than 50 lattice sites. The results also show a broadening of the wavepacket at each turning point of the super oscillation which leads to a dephasing of the SBO. This broadening is related to the abrupt change in amplitude and frequency of the SBO due to the spatial inhomogeneity. One can interpret this in terms of a modulated detuning that arises due to spatial variations of the nominal local Bloch frequency. For a fixed driving frequency, the spatially-varying Bloch frequency give rise to a detuning which increases for wavepacket transport against the direction of the force and which is reduced when the transport is in the direction of force. As the amplitude and the period of the SBO are directly related, i.e., inversely proportional, to the detuning they are continuously changing during the spatio-temporal evolution. Thus for transport across many lattice sites a significant wavepacket broadening emerges. This feature is controllable by an external detuning and the SBOs can be seen following a sawtooth oscillation profile.
\begin{figure}[t]
\includegraphics[scale=0.4]{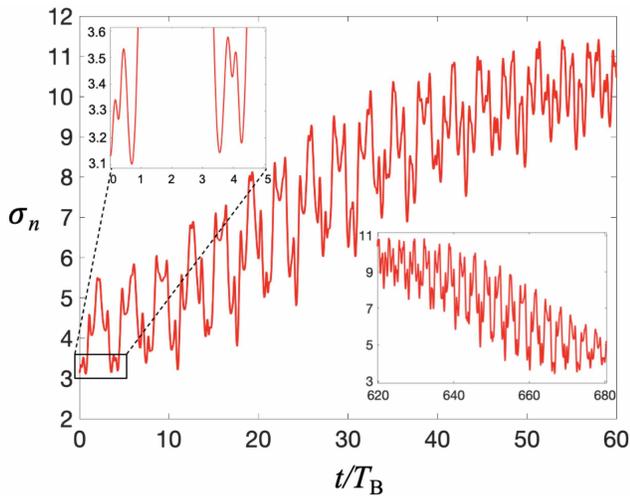}
\caption{\label{fig:epsart} Width of the wavepacket in real space for a drive phase of $\phi=0$. The left inset shows the magnified image of the width during the first five Bloch periods. The inset on the right displays the long time evolution of the width which shows a decline of the width corresponding to a complete revival. }
\end{figure}
\par{In Fig.~2(b) we show the SBOs dynamics in k-space which confirms that the spatial inhomogeneity leads to the existence of an effective relative phase between the modulation and the BOs in the shaken periodic lattice even without an external detuning.
The modulation is seen to act on top of the linear translation of the quasimomentum with time and the phase is such that the interaction between the modulation and the BOs takes place at quasimomentum values near the center of the BZ. This variable, i.e., space and thus time dependent, detuning provides rapid oscillations of the phase which, here, do not cover the entire BZ. Thus the superperiod $T_{SBO}$ is comprised of few Bloch periods $T_B$ and the amplitude is relatively low as compared to the case of a constant Bloch frequency with a static detuning. In Fig.~2(b) we see a clear manifestation of the acceleration theorem \cite{14,24}, which here is confirmed for the case of a non-constant detuning. Analyzing the dynamics in both real and k space provides a clear picture and shows that the spatial variations of Bloch frequency modify the propagation of the quasimomentum and the evolution of the quasimomentum can be used to understand the dynamics in real space.

\begin{figure}[t]
\includegraphics[scale=0.55]{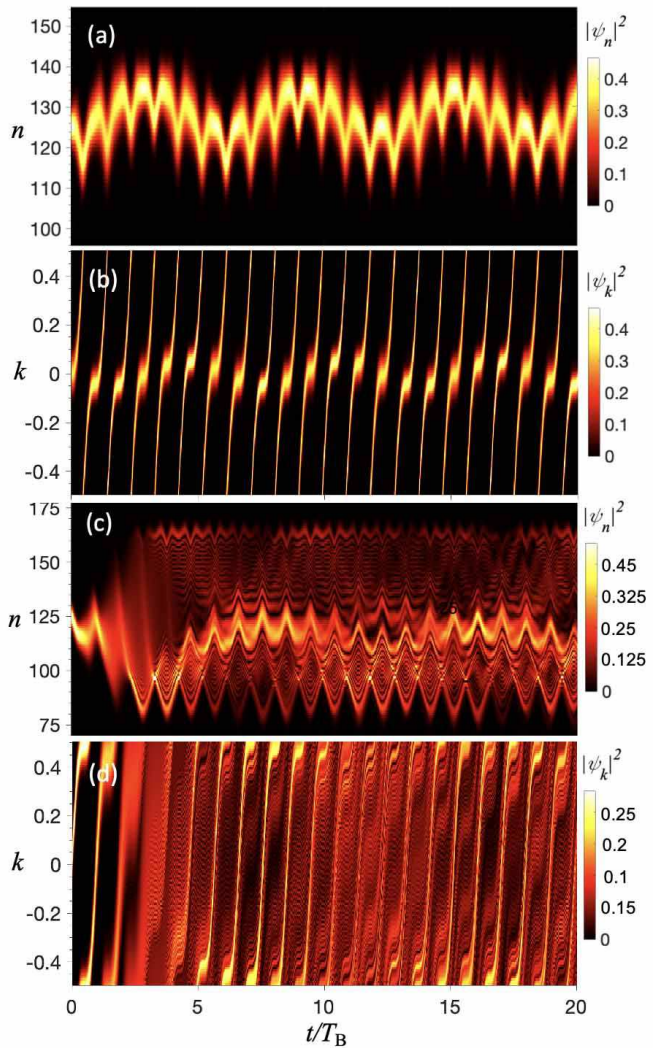}
\caption{\label{fig:epsart} Drive-phase-dependent dynamics. Time evolution of the absolute square of the real space wavefunction for a drive phase of (a) $\phi = -\pi/2$ and (c) $\phi = \pi/2$ exhibiting coherent SBOs and asymmetric spreading oscillations, respectively. The corresponding quasi-momentum evolutions are shown in (b) and (d), respectively.} 
\end{figure}

\par{The wavepacket evolution in k-space also show similar decay as already seen in real space, however, the width in k-space is inverse to the one in real space. A decay leads to the collapse of coherent oscillations and the long time dynamics shows dephased SBOs.  To analyze the decay, we plot the time evolution of the width in real space, i.e., the square root of the variance $\Delta n$, in Fig.~3. Starting with the first super-period, we note that the width increases when the wavepacket moves against the force and becomes smaller during the motion in the direction of the force. However, the initial width is never returned after the first Bloch period, see insets in Fig.~3, and there is an overall growth every super period. The width is reduced at the start of second super-oscillation, which again increases in addition to a growth factor from the previous oscillation. This pattern is repeated until the width saturates at a maximum value. The oscillation of the width continues in the saturation region until an interval of coherent dynamics reappears. 
The inset at the right bottom of Fig.~3 shows that for long times the wavepacket gets narrower and at time equal to $680\;T_B$ a revival occurs.} 

\par{As shown in Fig.~4(a), a dephasing of the wavepacket is not present for a drive phase of $\phi=-\pi/2$. In this case we obtain coherent SBOs, which remain intact even on long time scales, however, the amplitude is reduced due to large initial detuning. The vanishing of the dephasing can be accredited to the reduction in the amplitude and the correspondingly decreased spatially-varying effects. Fig.~4(b) shows the same behavior in k-space. The modulation appears immediately with a relative phase following a sine form, whose amplitude is quite low and is again confined near the center of the BZ. Thus the interaction between BO and the modulation takes place at small values of the quasimomentum and accordingly the real space transport is reduced.}

\par{At this point we note that the amplitude of the super oscillations can be enhanced for modulations that modify the BO dynamics near extreme values of the quasimomentum. In Fig.~4(c) we show the real space dynamics for such a case where we modulate the system with a phase of $\phi=\pi/2$. Clearly we see an increase in the amplitude of the wavepacket transport, however, we find a new kind of dynamics which correspond to a superposition of ordinary and super Bloch dynamics. In contrast with the coherent oscillations reported above, the wavepacket distribution now spreads rapidly. The wavepacket stretches in space and we can identify two regions of unequal densities, where at one end with higher density towards the direction of force the wavepacket performs purely breathing dynamics and at the other end with lower density it undergoes rapid super Bloch breathing. Both breathing oscillations occurring at a difference of $0.5T_B$ pass on the maximum density to a revival of the coherent BO in just 6 Bloch periods. Furthermore, we see all three types of oscillations happening at the same time with periodic changes in density.

\begin{figure}[t]
\includegraphics[scale=0.4]{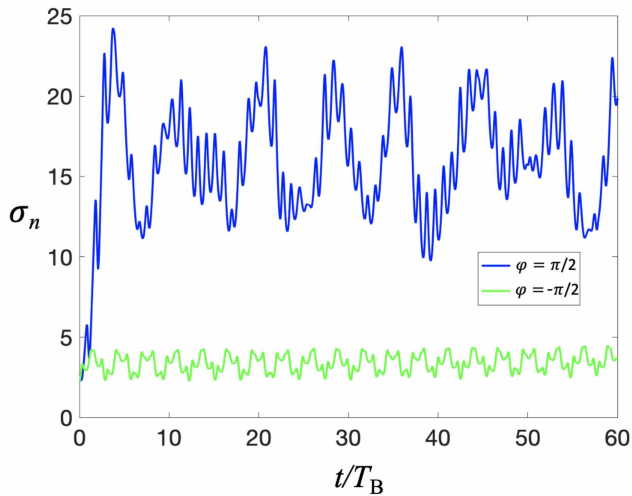}
\caption{\label{fig:epsart} Time dependence of the wavepacket width in real space for drive phases of $\phi=\pi/2$ (blue) and $\phi=-\pi/2$ (green).} 
\end{figure}

\par{The modulation appearing near the edges of the BZ is also responsible for the wavepacket spreading. This is evidenced by the wavepacket evolution in k-space. Fig.~4(d) highlights the sustained momenta near both edges which correspond to real space motion in opposite directions and by which wavepacket spreads at times where Bragg reflections occur.  One can also interpret this in terms of a negligible cycle-averaged momentum. We see that the average momentum slowly increases due to the varying Bloch frequency and thus an asymmetric spreading is generated. Then again the wavepacket gets narrower due to phase mixing and continues following a combination of breathing and coherent dynamics. We mention that the spreading oscillations reported here are similar in nature to the ballistic spreading regime present in the modulation of constant force system. However, under the same drive conditions we find an asymmetrical spreading motion which dies out in few Bloch periods, which unlike ballistic spreading gives rise to a mix of breathing and coherent Bloch dynamics.} 

\par{In Fig.~5 we present the width variations in real space for the wavepacket evolving in exactly opposite drive phases. These we compare with each other and to the width dynamics of a zero phase drive which we have discussed earlier. Here, the width in a $\phi= -\pi/2$ drive is seen to follow the super-Bloch oscillatory width profile of the $\phi= 0$ case. But now the width periodically returns to its initial value and there is no overall growth. The small width oscillations are sustained even at larger times and we perceive the case of $\phi= -\pi/2$ as a purely coherent regime of super-Bloch dynamics. On the contrary, the width increases sharply for the drive phase $\phi= \pi/2$, reaching a larger value which is much bigger than the maximum width reached in $\phi= 0$ dynamics. The width again saturates following even bigger width oscillations but it remains far above the initial  value and we have not seen a complete revival in the dynamics with $\phi= \pi/2$.}

\begin{figure}[t]
\hspace*{-0.3cm} 
\includegraphics[scale=0.48]{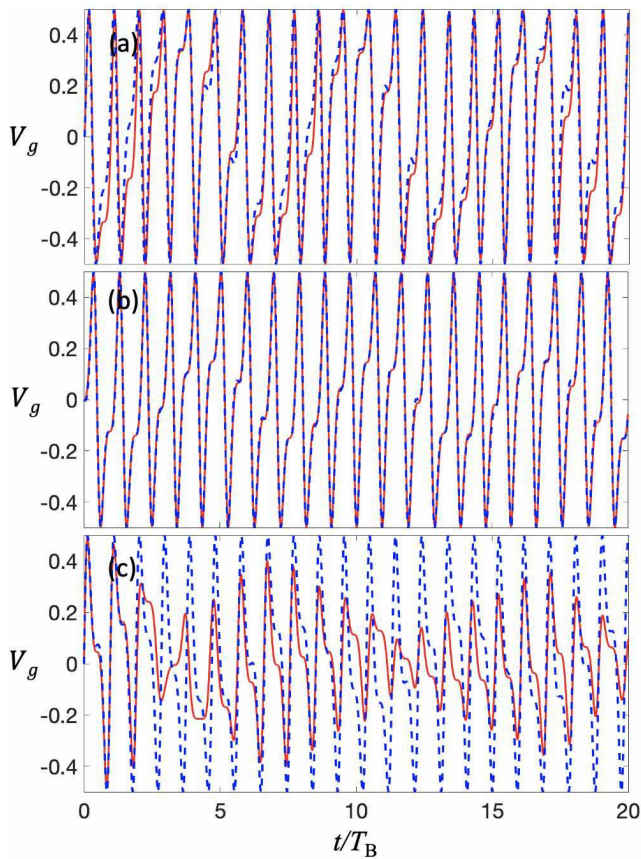}
\caption{\label{fig:epsart} Group velocity as a function of time for (a) $\phi=0$, (b) $\phi=-\pi/2$, and (c) $\phi=\pi/2$. The red line represents the result of numerical calculations, while the dashed blue line depict the dependence obtained from Eq.~(4).  For (a) the parametric values are $\Delta F= 3.65 \times 10^{-4}~E_R $ and $\hbar \delta \omega = 5.43\times 10^{-4}~E_R $, while in (b) and (c) $\Delta F= 2.13 \times 10^{-4}~E_R$. All the other parameters are the same as used previously.} 
\end{figure}

\par{We analyze the obtained complex dynamics further by comparing them to a simple analytical model. It is well known that the transport dynamics induced by the modulation of a constant force can be described by the semiclassical group velocity within a tight-binding model by considering the time-dependent quasi-momentum \cite{23,24,25}. With this starting point the super Bloch dynamics can be approximately captured by adding a constant detuning between the Bloch and the modulation frequencies. For SBOs appearing for the position-dependent force of parabolic trap, we can estimate the effect of the spatial variations of the Bloch frequency by the inclusion of a variable detuning. Following the slow oscillatory transport, the spatial variations are periodic and we approximate the detuning with a sinusoidal function, which also carries the initial detuning that depends on the drive phase. The group velocity is thus expressed as
\begin{eqnarray} 
	V_g(t) = \frac{Jd}{\hbar} \sin \!\bigg[k_o+ \frac{Fd}{\hbar} t - \frac{\Delta F d}{\hbar \delta\omega}\big\{\!\cos(\delta\omega t\!+\!\phi) \!-\!\cos(\phi) \big\} \nonumber \\ - \frac{F_Dd}{\hbar \omega_D} \;\!\big\{\cos(\omega_D t+\phi)-\cos(\phi)\big\} \bigg] , \quad\quad 
\end{eqnarray}
where $\Delta F$ is the amplitude of the force related to the oscillatory detuning and $\delta \omega$ is the frequency of the oscillation. Equation~(5) is plotted in Fig.~6 where the analytical result is compared with the numerically-calculated dynamics of the group velocity. The parameters $\Delta F$ and $\delta \omega$ are particular to the system and we extract these from the real space dynamics shown above.}

Figure~6(a) shows that for $\phi=0$ the semiclassical and approximate analytical result covers the relative phase of SBOs quite well. However, in this case the spatially-varying effects are quite significant and cannot be fully captured by the simple approximate model and therefore no exact match between numerics and Eq.~(5) is obtained. For the case of $\phi=-\pi/2$ shown in Fig.~6(b) we achieve a very good agreement between the Eq.~(5) and the numerical calculations and the rapid oscillations of the relative phase are also confirmed. Clearly, the insertion of an oscillatory detuning put restrictions on the relative phase such that the modulation now do not affect the entire velocity values unless $\delta \omega$ is very small. Fig.~6(c) shows the breakdown of our analytical model which is due to the spreading and the multi-mode dynamics present for $\phi=\pi/2$ which are beyond the simple semiclassical model.}

\section{Conclusions}

We demonstrate that the position- and time-dependent force realized by a modulated parabolic trap can be utilized to realize super-Bloch dynamics in the absence of an external detuning. Depending upon the initial phase of the modulation field the oscillations are generated with different amplitude. The small amplitude oscillations remain coherent, while at large amplitudes the oscillations decay, sometimes leading to dephasing and in other conditions producing new kind of mixed ordinary and super-Bloch dynamics. These different regimes depend on the drive phase and have been linked to the relative phase appearing in the dynamics in quasimomentum space. The good agreement with an approximate semiclassical model for some initial phases confirms our interpretations. In the future, we plan to work on developing improved analytical descriptions and will consider other implementations of the variable detuning.

\begin{acknowledgments}
The authors thank Martin Holthaus for helpful discussions. 
U.A. gratefully acknowledges support from the Deutscher Akademischer Austauschdienst (DAAD) by a doctoral research grant.
\end{acknowledgments}

\end{document}